# High-Uniformity Calculation Method of Four-Coil Configuration in Large-Caliber Magnetic Field Immunity Testing System

Xi Deng, Ya Huang, Chenguang Wan, Li Jiang*, Ge Gao, Zhengyi Huang and Jie Zhang

*Abstract*—**Power electronic equipment regulated by the International Thermonuclear Experimental Reactor (ITER) organization must pass the relevant steady-state magnetic field immunity test. The main body of magnetic field immunity test is magnetic field generator coil. Through mathematical derivation in this paper, the magnetic field calculation formulas of four-coil configuration under ideal and actual models are obtained. The traditional method of magnetic field performance calculation is compared with the general formula method under the ideal model. A global parameter optimization method based on Lagrange Multiplier by KKT conditions is proposed to obtain the coil parameters of high-uniformity magnetic field. The magnetic field distribution in the uniform zone is revealed by the finite element method. The model analysis is proved to be correct and effective by experimental results. The research of this paper provides a practical scheme for the coil design with high magnetic field and high-quality uniformity.**

## I. Introduction

ITER is the largest tokamak experimental reactor under construction in the world. The surrounding area will be covered by a high magnetic field during operation, which will affect the safety and reliability of various electronic and electrical equipment. The steady-state maximum magnetic field of the tokamak building outside the device can reach 200 mT. To ensure safe and stable operation, all power electronic equipment in ITER with a magnetic field greater than 5 mT in the surrounding environment of the hall must pass the relevant static magnetic field immunity test [1]. According to the distribution of the magnetic field required by ITER report [2], the test is divided into five levels. The nominal magnetic field of each test level and the requirements for testing the space magnetic field are shown in Table I. For each test level, to consider a certain margin, the magnetic field in the whole test space of the equipment under the test (EUT) is 1.4 times higher than the nominal magnetic field. Further, considering that high magnetic field may lead to damage of the EUT, it is required that the maximum magnetic field in the whole test space of the EUT should be twice the nominal magnetic field [3].

This work was supported in part by the National Key Research & Development Plan 2017YFE0300401 and Comprehensive Research Facility for Fusion Technology (No. 2018-000052-73-01-001228). (*Corresponding author: Li Jiang*)
X. Deng, C. Wan are with the Institute of Plasma Physics, Chinese Academy of Sciences, Hefei 230031, China, and also with the University of Science and Technology of China, Hefei 230026, China (e-mail: xi.deng@ipp.ac.cn; chenguang.wan@ipp.ac.cn).
Y. Huang, L. Jiang, G. Gao, Z. Huang, J. Zhang are with the Institute of Plasma Physics, Hefei Institutes of Physical Science, Chinese Academy of Sciences, Hefei 230031, China (e-mail: ya.huang@ipp.ac.cn; jiangli@ipp.ac.cn; gg@ipp.ac.cn; hzy@ipp.ac.cn; zhangjie@ipp.ac.cn).

TABLE I
TEST LEVEL OF THE STEADY MAGNETIC FIELD

| Test level | Nominal magnetic field | Min/Max magnetic field |
|---|---|---|
| 1 | 7.5 mT | 10.5 mT/15 mT |
| 2 | 15 mT | 21 mT/30 mT |
| 3 | 30 mT | 42 mT/60 mT |
| 4 | 60 mT | 84 mT/120 mT |
| 5 | 120 mT | 168 mT/240 mT |
| 6 | $n$ | 1.4 $n$ / 2 $n$ |

The following characterization parameters can be introduced in the uniform zone:
1) Side length of square uniform zone (2$s$).
2) Minimum magnetic field ($B_{min}$): Minimum value of magnetic field space in the uniform zone.
3) Maximum magnetic field ($B_{max}$): Maximum value of magnetic field space in the uniform zone.
4) Magnetic field uniformity ($\eta$): Ratio of maximum to minimum magnetic field.

$$\eta = \frac{B_{max}}{B_{min}} \quad (1)$$

There are many devices for generating uniform magnetic field, including solenoid coil, Helmholtz coil, multi-coil group, three-dimensional orthogonal coil, etc. Solenoid coil is a common structure, scientists all over the world have carried out detailed research on the structure, principle, test and application of the coil [4]-[9]. Helmholtz coil is the most studied coil structure with simple structure. The configuration, principle, and optimization of a Helmholtz coil are described and analyzed in detail in [10]-[18]. For applications that require a higher magnetic field uniformity, the multi-coil configuration can be used to solve the problem. Scientists have studied the magnetic field uniformity of different coil groups and provided reference data in various cases [19]-[31]. The parameters of the multi-coil group configuration are optimized and analyzed [31], and the optimal parameters under each structure are obtained. Orthogonal coil configuration also has corresponding references for research and analysis [32] [33].

Because square structure has more advantages than circular structure in the manufacturing and welding process of large equipment [34] [35], it is recommended by ITER organization [1] and IEC (International Electrotechnical Commission) [36]-[38] in magnetic field immunity test. For the magnetic field immunity testing equipment with test level 4, $B_{max}$=120 mT and $\eta \leq 1.2$, the design, manufacture and test have been completed [3][39]. However, the equipment cannot provide help for the requirements of stronger magnetic field and higher uniformity (Test level 5, $B_{nom}$=120 mT and $\eta \leq 1.05$). On this basis, this paper designs the equipment with higher requirements ($B_{max}$=275 mT and $\eta \leq 1.05$).

This paper is organized as follows. In Section II, the calculation formulas of uniform regional magnetic field under ideal model and actual model are derived. In Section III, the traditional methods of magnetic field performance calculation and the general formula under the ideal model are compared. Based on Lagrange Multiplier Method by KKT conditions, the global parameter optimization is carried out with the actual model to obtain the optimal scheme, and the magnetic field is analyzed by the finite element method. In Section IV, experimental verification is provided. The conclusion is given in Sections V.

## II. MAGNETIC FIELD CALCULATION OF FOUR-COIL CONFIGURATION

### A. Ideal model

The four-coil ideal configuration is shown in Fig. 1. The side length of four coils is $2l$, the distance between two inner coils is $2h_1$, the distance between two outer coils is $2h_2$, the ampere turns of two inner coils are $N_1I$, and the ampere turns of two outer coils are $N_2I$.

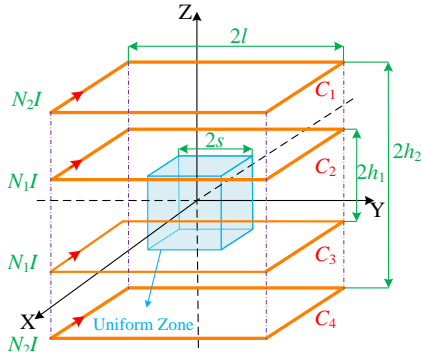

Fig. 1. Four-coil ideal configuration model.

According to Biot-Savart's Law and the magnetic field superposition principle, the magnetic field $B_I(x, y, z)$ produced by the square four-coil ideal configuration at the point $P(x, y, z)$ in the uniform zone can be expressed as

$$\begin{cases} B_I(x) = \dfrac{\mu_0 I}{4\pi} \cdot \sum_{t=1}^{2}\left(N_t \cdot \sum_{i=0}^{1}\sum_{j=0}^{1}\sum_{k=0}^{1}\dfrac{(-1)^{i+1} Y_1 \cdot Z_t}{X_1^2 + Z_t^2} \cdot D_t\right) \\ B_I(y) = \dfrac{\mu_0 I}{4\pi} \cdot \sum_{t=1}^{2}\left(N_t \cdot \sum_{i=0}^{1}\sum_{j=0}^{1}\sum_{k=0}^{1}\dfrac{(-1)^{j+1} X_1 \cdot Z_t}{Y_1^2 + Z_t^2} \cdot D_t\right) \\ B_I(z) = \dfrac{\mu_0 I}{4\pi} \cdot \sum_{t=1}^{2}\left(N_t \cdot \sum_{i=0}^{1}\sum_{j=0}^{1}\sum_{k=0}^{1}\left[\dfrac{X_1 \cdot Y_1}{X_1^2 + Z_t^2} + \dfrac{X_1 \cdot Y_1}{Y_1^2 + Z_t^2}\right] \cdot D_t\right) \\ B_I(x,y,z) = \left(B_I^2(x) + B_I^2(y) + B_I^2(z)\right)^{1/2} \end{cases} \quad (2)$$

where $|x| \leq s$, $|y| \leq s$, $|z| \leq s$, $X_1=(-1)^i x+l$, $Y_1=(-1)^j y+l$, $Z_t=(-1)^k z+h_t$, $D_t=(X_1^2+Y_1^2+Z_t^2)^{1/2}$.

The turn ratio of the side coil to the intermediate coil is defined as $\beta$ and can be described as:

$$\beta = N_2 / N_1 \quad (3)$$

The magnetic field value and uniformity in the uniform zone are functions of parameters $l$, $h_1$, $h_2$, $\beta$, $s$ and $N_1I$.

$$\begin{cases} B_{max} = f_1(x,y,z|l,h_1,h_2,\beta,s) \cdot N_1 I \\ B_{min} = f_2(x,y,z|l,h_1,h_2,\beta,s) \cdot N_1 I \\ \eta = g(x,y,z|l,h_1,h_2,\beta,s) \end{cases} \quad (4)$$

### B. Actual model

Compared with the ideal model, the actual model has the problems of spiral structure and conductor section.

#### 1) Helical structure

The helical structure is analyzed from one turn coil. The one turn coil is shown in Fig. 2. Where $\theta$ is the offset angle of one side length, $2l$ is the side length of the primary coil, and its relationship is

$$\theta = \arctan \dfrac{d}{8l} \quad (5)$$

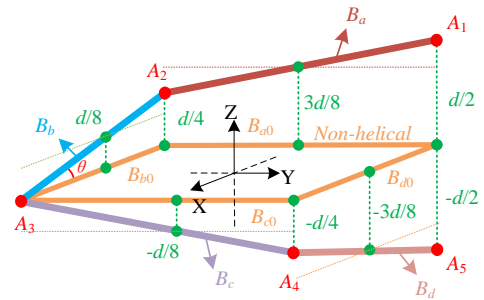

Fig. 2. One turn coil for helical structure.

The magnetic field generated by non-helical one turn coil at the point $P(x, y, z)$ in the uniform zone is

$$\begin{cases} B_{a0}(x) = \dfrac{\mu_0 I}{4\pi s} \cdot \dfrac{-z}{X_0^2 + z^2}\left(\dfrac{Y_1}{D_{01}} + \dfrac{Y_0}{D_{00}}\right) \\ B_{a0}(y) = 0 \\ B_{a0}(z) = \dfrac{\mu_0 I}{4\pi s} \cdot \dfrac{X_0}{X_0^2 + z^2}\left(\dfrac{Y_1}{D_{01}} + \dfrac{Y_0}{D_{00}}\right) \end{cases} \quad (6)$$

$$\begin{cases} B_{b0}(x) = 0 \\ B_{b0}(y) = \dfrac{\mu_0 I}{4\pi s} \cdot \dfrac{-z}{Y_0^2 + z^2}\left(\dfrac{X_0}{D_{00}} + \dfrac{X_1}{D_{10}}\right) \\ B_{b0}(z) = \dfrac{\mu_0 I}{4\pi s} \cdot \dfrac{Y_0}{Y_0^2 + z^2}\left(\dfrac{X_0}{D_{00}} + \dfrac{X_1}{D_{10}}\right) \end{cases} \quad (7)$$

$$\begin{cases} B_{c0}(x) = \dfrac{\mu_0 I}{4\pi s} \cdot \dfrac{z}{X_1^2 + z^2}\left(\dfrac{Y_1}{D_{10}} + \dfrac{Y_0}{D_{11}}\right) \\ B_{c0}(y) = 0 \\ B_{c0}(z) = \dfrac{\mu_0 I}{4\pi s} \cdot \dfrac{X_1}{X_1^2 + z^2}\left(\dfrac{Y_1}{D_{10}} + \dfrac{Y_0}{D_{11}}\right) \end{cases} \quad (8)$$

$$\begin{cases} B_{d0}(x) = 0 \\ B_{d0}(y) = \dfrac{\mu_0 I}{4\pi s} \cdot \dfrac{z}{Y_1^2 + z^2}\left(\dfrac{X_1}{D_{11}} + \dfrac{X_0}{D_{01}}\right) \\ B_{d0}(z) = \dfrac{\mu_0 I}{4\pi s} \cdot \dfrac{Y_1}{Y_1^2 + z^2}\left(\dfrac{X_1}{D_{11}} + \dfrac{X_0}{D_{01}}\right) \end{cases} \quad (9)$$

where $X_i=(-1)^i x+l$, $Y_j=(-1)^j y+l$, $D_{ij}=(X_i^2+Y_j^2+z^2)^{1/2}$. $i, j=\{0,1\}$.

On the basis of the original conductor ($B_{a0}$), translate the distance of $3d/8$ along the positive direction of the Z axis, and then rotate the angle of $\theta$ along the YZ direction around the X axis to obtain the conductor $B_a$. The conductor $B_b$, $B_c$ and $B_d$ are obtained by the same conversion method.

The magnetic field generated by the non-helical conductor is obtained by the coordinate transformation of translation and rotation, and then decomposed into the X, Y and Z axis components of the original coordinate system.

$$\begin{cases} [B_a(x)\ B_a(y)\ B_a(z)]^T = A[B_{a0}(x_a)\ B_{a0}(y_a)\ B_{a0}(z_a)]^T \\ [B_b(x)\ B_b(y)\ B_b(z)]^T = B[B_{b0}(x_b)\ B_{b0}(y_b)\ B_{b0}(z_b)]^T \\ [B_c(x)\ B_c(y)\ B_c(z)]^T = C[B_{c0}(x_c)\ B_{c0}(y_c)\ B_{c0}(z_c)]^T \\ [B_d(x)\ B_d(y)\ B_d(z)]^T = D[B_{d0}(x_d)\ B_{d0}(y_d)\ B_{d0}(z_d)]^T \end{cases} \quad (10)$$

$$\begin{cases} [x_a\ y_a\ z_a]^T = C[x\ y\ z]^T + [0\ 0\ 3d/8]^T \\ [x_b\ y_b\ z_b]^T = D[x\ y\ z]^T + [0\ 0\ d/8]^T \\ [x_c\ y_c\ z_c]^T = A[x\ y\ z]^T + [0\ 0\ -d/8]^T \\ [x_d\ y_d\ z_d]^T = B[x\ y\ z]^T + [0\ 0\ -3d/8]^T \end{cases} \quad (11)$$

where $A = \begin{bmatrix} 1 & 0 & 0 \\ 0 & \cos\theta & -\sin\theta \\ 0 & \sin\theta & \cos\theta \end{bmatrix}$, $B = \begin{bmatrix} 0 & \cos\theta & \sin\theta \\ 1 & 0 & 0 \\ 0 & -\sin\theta & \cos\theta \end{bmatrix}$,

$C = \begin{bmatrix} 1 & 0 & 0 \\ 0 & \cos\theta & \sin\theta \\ 0 & -\sin\theta & \cos\theta \end{bmatrix}$, $D = \begin{bmatrix} 0 & \cos\theta & -\sin\theta \\ 1 & 0 & 0 \\ 0 & \sin\theta & \cos\theta \end{bmatrix}$.

According to the magnetic field superposition principle, the magnetic field generated by the point $P(x, y, z)$ in the uniform zone of the intermediate one turn coil is

$$\begin{cases} B_0(x) = B_a(x) + B_b(x) + B_c(x) + B_d(x) \\ B_0(y) = B_a(y) + B_b(y) + B_c(y) + B_d(y) \\ B_0(z) = B_a(z) + B_b(z) + B_c(z) + B_d(z) \end{cases} \quad (12)$$

For the $C_1$ coil in Fig. 1, the location diagram is shown in Fig. 3, and there are $N_2$ turns in total. The magnetic field generated by the coil with other turns is equivalent to the translation on the Z axis. The transformation formula is

$$[x_{r_1}\ y_{r_1}\ z_{r_1}]^T = [x\ y\ z]^T + \left[0\ 0\ \left(h_2 - \dfrac{d(N_2-1)}{2} + r_1 \cdot d\right)\right]^T \quad (13)$$

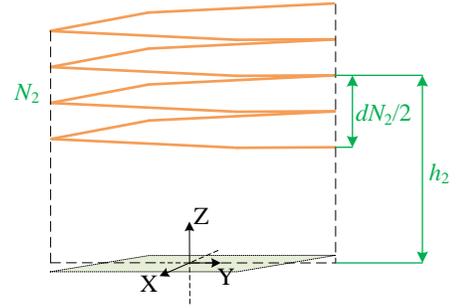

Fig. 3. Location diagram of $C_1$ coil.

The magnetic field generated by $N_2$ turns in the $C_1$ is

$$\begin{cases} B_{C1}(x) = \sum_{r_1=0}^{N_2} B_0(x_{r_1}) \\ B_{C1}(y) = \sum_{r_1=0}^{N_2} B_0(y_{r_1}) \\ B_{C1}(z) = \sum_{r_1=0}^{N_2} B_0(z_{r_1}) \end{cases} \quad (14)$$

Similarly, the magnetic field generated by other coil groups of four-coil configuration can be obtained, and the total magnetic field $B_H(x, y, z)$ is obtained by superposition.

$$\begin{cases} B_H(x) = B_{C1}(x) + B_{C2}(x) + B_{C3}(x) + B_{C4}(x) \\ B_H(y) = B_{C1}(y) + B_{C2}(y) + B_{C3}(y) + B_{C4}(y) \\ B_H(z) = B_{C1}(z) + B_{C2}(z) + B_{C3}(z) + B_{C4}(z) \\ B_H(x, y, z) = (B_H^2(x) + B_H^2(y) + B_H^2(z))^{1/2} \end{cases} \quad (15)$$

### 2) Cross-section

The shape of the coil conductor cross-section has a certain impact on the magnetic field performance [17] [18]. The cross-section model is shown in Fig. 4.

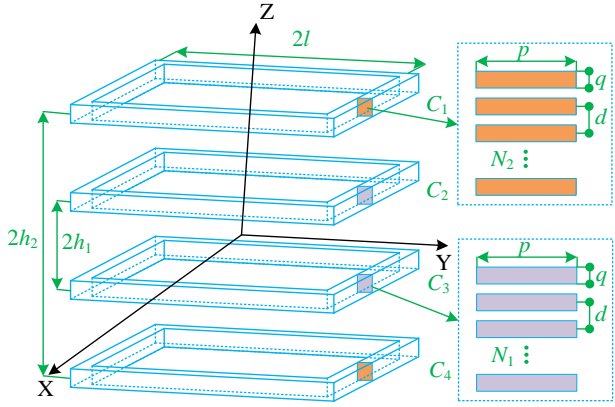

Fig. 4. The cross-section models.

The cross-section length $L$ of the four-coil groups is $[L_1, L_2]$, $L_1= l-p/2$, $L_2= l+p/2$.

The height $H_{1r}$ of $C_1$ coil is $[H_1+r_1d, H_1+q+r_1d]$, $H_1=h_2-(d\cdot(N_2-1)-q)/2$, $r_1= \{0, 1\ldots, N_2\}$.

The height $H_{2r}$ of $C_2$ coil is $[H_2+r_2d, H_2+q+r_2d]$, $H_2=h_1-(d\cdot(N_1-1)-q)/2$, $r_2= \{0, 1\ldots, N_1\}$.

The height $H_{3r}$ of $C_3$ coil is $[H_3+r_3d, H_3+q+r_3d]$, $H_3=-h_1-(d\cdot(N_1-1)-q)/2$, $r_3= \{0, 1\ldots, N_1\}$.

The height $H_{4r}$ of $C_4$ coil is $[H_4+r_4d, H_4+q+r_4d]$, $H_4=-h_2-(d\cdot(N_2-1)-q)/2$, $r_4= \{0, 1\ldots, N_2\}$.

The magnetic field distribution in the uniform zone can be expressed as

$$B_{pq} = \frac{1}{pq}\left(\sum_{t=1}^{3}\left(\begin{array}{l}\sum_{r_1=0}^{N_2}\left(\int_{L_1}^{L_2}\int_{H_1+r_1d}^{H_1+r_1d+q}B_{C1}(u_t|L,H_{1r})dLdH_{1r}\right)+\\ \sum_{r_2=0}^{N_1}\left(\int_{L_1}^{L_2}\int_{H_2+r_2d}^{H_2+r_2d+q}B_{C2}(u_t|L,H_{2r})dLdH_{2r}\right)+\\ \sum_{r_3=0}^{N_1}\left(\int_{L_1}^{L_2}\int_{H_3+r_3d}^{H_3+r_3d+q}B_{C3}(u_t|L,H_{3r})dLdH_{3r}\right)+\\ \sum_{r_4=0}^{N_2}\left(\int_{L_1}^{L_2}\int_{H_4+r_4d}^{H_4+r_4d+q}B_{C4}(u_t|L,H_{4r})dLdH_{4r}\right)\end{array}\right)^2\right)^{\frac{1}{2}} \quad (16)$$

where $u_1=x$, $u_2=y$, $u_3=z$.

Comprehensive consideration shows that the magnetic field performance is related to parameters such as $l$, $h_1$, $h_2$, $d$, $p$, $q$, $s$, $N_1I$ and $N_2I$, which can be expressed as

$$\begin{cases}B_{\max} = f_1(x,y,z|l,h_1,h_2,d,p,q,s,N_1,N_2)\cdot I\\ B_{\min} = f_2(x,y,z|l,h_1,h_2,d,p,q,s,N_1,N_2)\cdot I\\ \eta = g(x,y,z|l,h_1,h_2,d,p,q,s,N_1,N_2)\end{cases} \quad (17)$$

## III. RESEARCH OF THE HIGH-UNIFORMITY MAGNETIC COIL DESIGN METHOD

### A. Traditional Method

The traditional method is to use Taylor expansion to calculate the coil structure parameters when the higher order of the axis is zero [17][35][40]. The axial magnetic field is equal, and the calculation is relatively simple, but it is not optimal for the case that the uniform zone is a cube.

$$B(z) = B_z(0)+\sum_{k=1}^{\infty}B_z^{2k}(0)\frac{z^{2k}}{(2k)!} \quad (18)$$

The parameter relationship of Merritt four-coil configuration is obtained as [35]

$$\begin{cases}h_1 = 0.256212\cdot l\\ h_2 = 1.010984\cdot l\\ \beta = 2.361197\\ N_2I=1.1142\cdot l\cdot B(0)\times 10^6\end{cases} \quad (19)$$

### B. General formula method

For the ideal model, the optimized general formulas of four-coil structure and magnetic field performance parameters can be summarized [30]

$$\begin{cases}l/s = 1.267+137.6\cdot e^{-5.874\eta}\\ h_1/s = 0.1616\cdot l/s - 0.1594\\ h_2/s = 1.1942\cdot l/s - 0.3880\\ \beta = 0.9779+14.19\cdot e^{-2.375\eta}\\ N_1I+N_2I=(1.964+458.9\cdot e^{-6.682\eta})sB_{\min}\times 10^6\end{cases} \quad (20)$$

The calculation of general formula is convenient and fast, but there are still several problems:

(1) The calculation range of magnetic field uniformity is $1.1\leq\eta\leq 1.4$, which cannot meet the requirements of high-uniformity ($\eta\leq 1.05$) magnetic field performance.

(2) The influence of helical structure and conductor cross-section are not considered in the calculation.

(3) Only the axial magnetic field is considered in the calculation, and the magnetic field of the actual model exists on the three axes (X, Y, Z), without considering the proportion of the magnetic field on each axis.

### C. Basic principle of Lagrange Multiplier Method by KKT conditions

This paper mainly considers the optimization of multi parameters. The generalized Lagrange multiplier method by KKT conditions is a good choice for it.

The Lagrange multiplier method is widely used for solving constrained optimization problems. For optimization problems with equality constraints, the Lagrange multiplier method can be used to find the optimal value; if there are inequality constraints, the KKT (Karush-Kuhn-Tucker) conditions can be used to generalize the Lagrange multiplier method to find the optimal value. Generally, the optimization problem with inequality constraints is solved as follows.

$$\begin{cases}\min f(x)\\ s.t. \ h_i(x) = 0, (i = 1,2,\ldots m)\\ \quad\quad g_j(x) \leq 0, (j = 1,2,\ldots n)\end{cases} \quad (21)$$

where $f(x)$ is the original objective function. $h_i(x)$, $g_j(x)$ is equality and inequality constraint function. $x= [x_1, x_2, x_3\ldots]^T$ is independent variable.

The Lagrange function is defined as:

$$L(x,\lambda,\mu) = f(x) + \sum_{i=1}^{m}\lambda_i h_i(x) + \sum_{j=1}^{n}\mu_j g_j(x) \quad (22)$$

where $\lambda = [\lambda_1, \lambda_2, \lambda_3..., \lambda_i]^T$, $\mu = [\mu_1, \mu_2, \mu_3..., \mu_j]^T$ is the Lagrange multiplier of $h_i(x)$, $g_j(x)$, respectively.

For the Lagrange multiplier method under simultaneous constraints of inequality and equality, the condition of the optimal solution can be expressed by the following KKT conditions:

$$\begin{cases} \nabla_x L(x^*,\lambda,\mu) = 0 \\ h_i(x^*) = 0 \\ g_j(x^*) \leq 0 \\ \mu_j > 0 \\ \mu_j \cdot g_j(x^*) = 0 \\ \nabla_\lambda L(x^*,\lambda,\mu) = 0 \\ \nabla_\mu L(x^*,\lambda,\mu) = 0 \end{cases} \quad (23)$$

where $x^*$ is extreme point of Eq.(21).

### D. Global optimization method---Lagrange Multiplier Method by KKT conditions

The optimization model is constructed as follows.

**Part I:** The optimization objectives are described.

The optimal state is the minimum weight and power loss. The total weight $G$ and power loss $P$ can be expressed as

$$\begin{cases} G = \rho V = 16\dfrac{\rho}{J} \cdot (N_1 + N_2) \cdot I \cdot l \\ P = \dfrac{J^2 V}{\sigma} = 16\dfrac{J}{\sigma} \cdot (N_1 + N_2) \cdot I \cdot l \end{cases} \quad (24)$$

where $\rho$ is the conductor mass density, $J$ is the conductor current density, $\sigma$ is the conductor conductivity, $V$ is the total volume.

When the parameter $A=(N_1+N_2)\cdot I\cdot l$ is the smallest, the weight and power loss are the smallest, which can be used as the optimization goal.

**Part II:** The equality and inequality constraints are considered.

According to the design requirement, the magnetic field $B_{max}$=275mT. The magnetic field uniformity $\eta$ is 1.05. The size of the test zone is 1×1×1m. A space of 0.325 m is reserved at each side for the installation of the equipment under test (EUT) handling platform, thus the inner space ($a=2\times(l$-$p/2)$) of the test coil should be 1.65 m at least. The separation distance of the inner two coils ($d_2=2\times H_2$) is 0.2m. The separation distance of the inner coil and the outer coil ($d_1= H_1-(H_2+q+r_2d)$) is in the range of 120mm to 200mm. The spacing $d$ between each turn of the coil should be less than 10mm. The current density cannot exceed 2.5A/mm$^2$. Limited by the output power of the power supply, the current per turn of the coil shall not exceed 12.5kA.

Thus, the optimization model is shown as follows:

$$\min A = (N_1 + N_2) \cdot I \cdot l$$

$$s.t. \begin{cases} s = 1m \\ d_2 = 200mm \\ \eta \leq 1.05 \\ I \leq 12.5kA \\ d \leq 10mm \\ 120mm \leq d_1 \leq 200mm \\ J \leq 2.5 A/mm^2 \\ a = 2\times(l-p/2) \geq 1650mm \\ B_{max} = 275mT \end{cases} \quad (25)$$

where

$$\begin{cases} B_{max} = f_1(x,y,z|l,d_1,d_2,p,q,d,s,N_1,N_2) \cdot I \\ B_{min} = f_2(x,y,z|l,d_1,d_2,p,q,d,s,N_1,N_2) \cdot I \\ \eta = g(x,y,z|l,d_1,d_2,p,q,d,s,N_1,N_2) \end{cases} \quad (26)$$

Based on the Eq.(5)-(17) and (21)-(26), one certain group of optimized coil parameters with the minimum value of $A=(N_1+N_2)\cdot I\cdot l$ can be obtained for a specific homogeneous field. Since the model is a symmetrical structure, only 1/8 of the structure is calculated.

After applied the KKT, we are given a normal optimization problem. A descent method was generally used for normal optimization problem consisting of the following steps.

a) Choose an initial feasible solution $x=[x_1, x_2, x_3...]^T$. Here, we initialize the parameters using the result of Taylor expansion in Eq.(18) and (19) and previous engineering experience.
b) Identify a feasible "target" solution $x^D$ in a "downhill direction." with momentum term.
c) Choose a step size $\alpha=0.01$, and set $x=\alpha x^D+(1-\alpha)x$.
d) Test for termination, and return to step (b) if we need to improve further.

After the above steps, we can get the optimal value. The calculation result is as follows. $I$=12.23kA, $N_1$=19.82, $N_2$=12.11, $l$=1093mm, $d_1$=140.80mm, $d_2$=200.55mm, $d$=8mm, $p$=268.8mm, $q$=27.2mm. Since the number of turns is an integer, $N_1$ and is selected as 20 and 12, and considering the accuracy of engineering fabrication, the values of other parameters are as Table II.

TABLE II
PARAMETERS OF THE FOUR-COIL SYSTEM

| Parameter | $\eta$ | $B_{min}$ | $B_{max}$ | $I$ | $N_1$ | $N_2$ | $s$ |
|---|---|---|---|---|---|---|---|
| Value | 1.0495 | 262.03 | 275 | 12.23 | 20 | 12 | 1000 |
| Unit | -- | mT | mT | kA | -- | -- | mm |
| Parameter | $2l$ | $d_1$ | $d_2$ | $d$ | $p$ | $q$ | |
| Value | 2190 | 140 | 200 | 8 | 270 | 27 | |
| Unit | mm | mm | mm | mm | mm | mm | |

### E. Calculation of FEM

In this article, finite-element software is used for the simulation calculation. The simulation model of the four-coil system is under an actual state, where the conductor cross section size and the helical structure is considered. $\eta$=1.05 and

$B_{min}$=275 mT is taken as an example. According to the above design parameters, the simulation model and results of the actual model are obtained, as shown in Fig. 5.

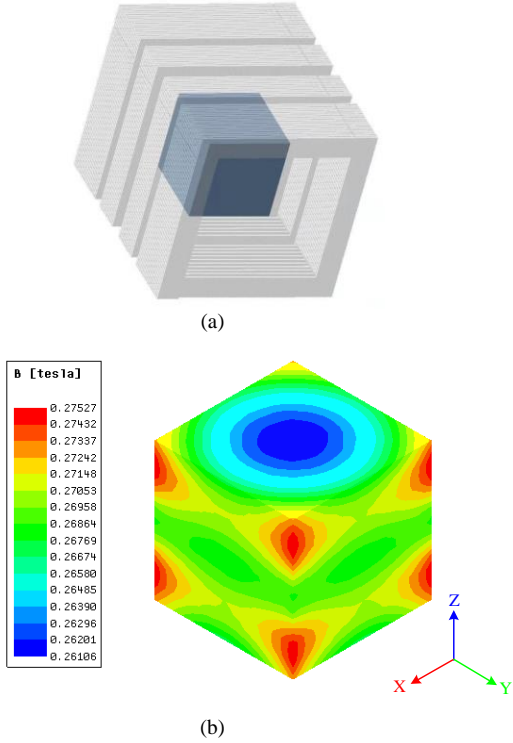

(a)

(b)

Fig. 5. Simulation model and results of the actual model: (a) Simulation model, (b) Distribution of the magnetic-field after optimization.

TABLE III
ERROR COMPARISON BETWEEN THE ANALYTIC METHOD AND THE FEM

| Structure | $B_{max}$/mT | $B_{min}$/mT | $\eta$ |
|---|---|---|---|
| Analytic Method | 275.00 | 262.03 | 1.0495 |
| FEM | 275.27 | 261.06 | 1.054 |
| Error | -0.098% | 0.370% | -0.428% |

The data from the finite-element method (FEM) simulation is compared with the analytical method calculation, as shown in Table III. The calculation error is less than 0.428%. The final simulation result is $\eta$ = 1.054. Although the value is slightly larger than 1.05 of the design requirements, it is still within 1.1 of the standard requirements. The design meets the requirements.

The magnetic field distribution of each surface in the optimized test area is intercepted. The intercepted surfaces are shown in Fig. 6, which are $z = 0$, $z = 0.5s$, $z = s$, $y = 0$, $y = 0.5s$ and $y = s$, respectively. The x-axis and y-axis are symmetrical in terms of magnetic field, and only one of them need to be considered.

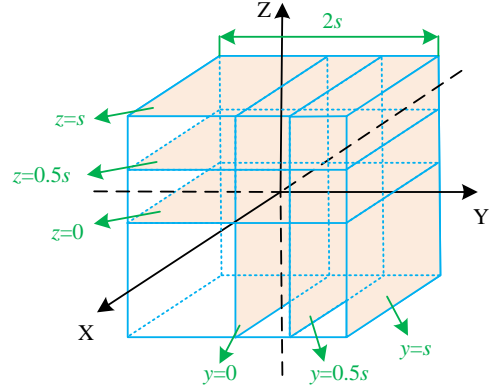

Fig. 6. Cut surface distribution of test area

Through FEM simulation, the magnetic field distribution on the six surfaces is calculated, as shown in Fig. 7. It can be seen that the magnetic field distribution inside the test area, in which the closer to the $z = 0$ plane in the axial direction ($z$-axis direction), the more uniform the magnetic field distribution. Compared with the $z$-plane, the magnetic field in the $y$-plane changes more violently, and the magnetic field fluctuation is the largest on the $y = s$ plane.

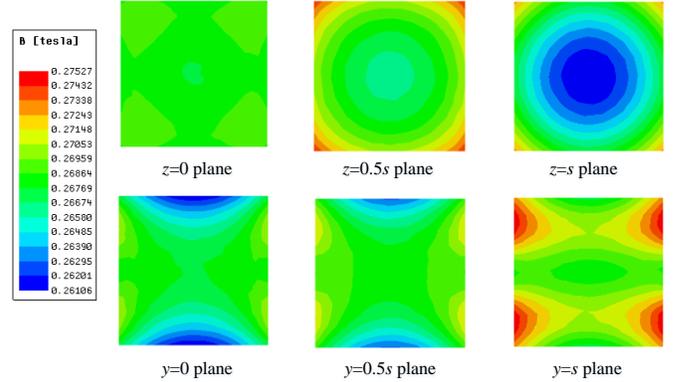

Fig. 7. Magnetic field distribution on different surfaces of the test area after optimization.

Similarly, calculate the magnetic field distribution on six line segments ($x=0,y=0$),($x=0,y=0.5s$),($x=0.5s,y=0.5s$),($x=0,y=s$),($x=0.5s,y=s$),($x=s,y=s$) in the z-axis direction and six line segments ($x=0,z=0$),($x=0,z=0.5s$),($x=0.5s,z=0.5s$),($x=0,z=s$),($x=0.5s,z=s$), ($x=s,z=s$) in the y-axis direction, as shown in Fig. 8. It can be seen that the magnetic field distribution of several special lines in the test area of the optimized four coil groups. In the z-axis direction, the closer to the outer side, the greater the magnetic field fluctuation, and each line is almost close to a point ($z = 0$), indicating that the magnetic field uniformity of the $z$-plane where this point is located is high; In the y-axis direction, the fluctuation of each line itself is not large, but the magnetic field values between lines are quite different.

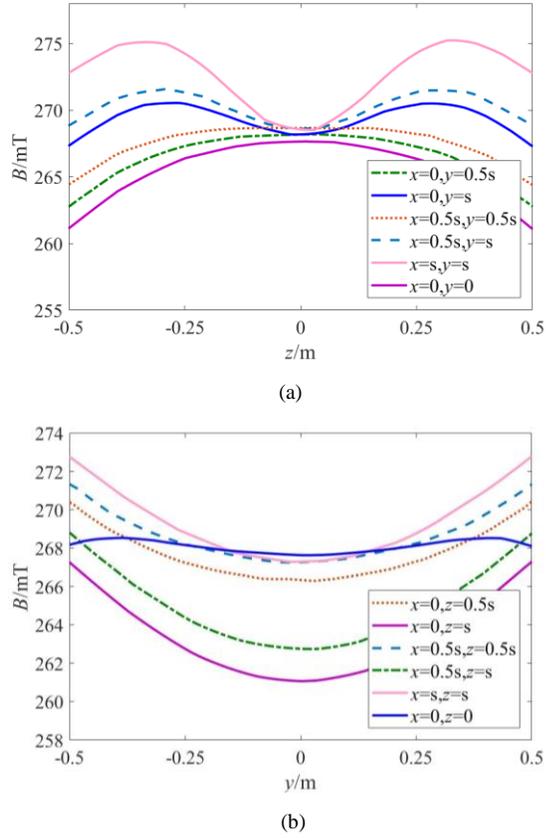

Fig. 8. The magnetic field distribution of different line segments in the test area, after optimization: (a) z-axis direction, (b) y-axis direction.

## IV. VERIFICATION

### A. Experimental Test

To verify the field inhomogeneity of the magnetic field generated by the test coil, the first step is to position the test zone. A frame structure should be manufactured first to install the probes. 8 probes (No. 3 ~ No. 10 in Fig. 9) located at the corners are used to position the test zone. If a coordinate system is defined at the center of the test zone as shown in Fig. 11O, the coordinates of these two points will be (0, 0, -0.5) and (0, 0, 0.5). As the maximum magnetic field occurs on the side edges of the test zone and its position is (0.5, 0.5, 0.32), 7 probes (No. 10 ~ No. 16) with a separation distance of 83.3 mm are used to measure the field distribution on a half of the edge. To measure the correct maximum value, the position of probe No. 12 can be adjusted to (0.5, 0.5, 0.32). The maximum field value will be obtained, and these values will be compared with the calculated ones to verify the theoretical results. It should be pointed out that, all the probes are used to measure the axial magnetic field (Z direction in Fig. 9).

Based on the analysis presented above, 16 Hall probes are employed to calibrate the test field. The probes can be divided into 3 groups, which are:

(1) 2 probes (No. 1 and No. 2) to measure the minimum magnetic field.

(2) 8 probes (No. 3 ~ No. 10) to locate the test zone.

(3) 6 probes (No. 11 ~ No. 16) along with No. 10 to measure the field distribution on the edge and the maximum magnetic field.

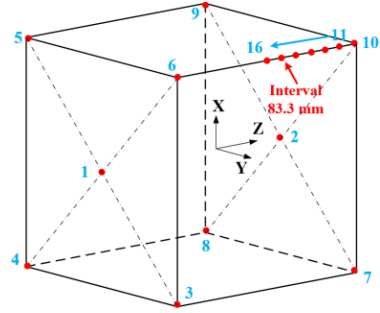

Fig. 9. Layout diagram of magnetic field measuring probe in uniform zone.

The testing model of the four-coil system and the testing system experimental platform is shown in Fig. 10. The system is insulated and fixed by the epoxy board, the water-cooled runner at the coil corner is connected by a water pipe, and the bottom is supported by the bottom foot. The deionized water is flown into each of the turn coils and leads at 1 m/s. The installation position accuracy of the Hall probe shall be within 1.5 mm from the designed locations.

When 12.23 kA current is applied, the magnetic-flux density measured by 16 Hall probes is compared with the simulation values, as shown in Table IV and Table V. Table IV shows that the errors are within 2% (standard error is less than 2%) and meet the requirements. The error comparison of $B_{max}$, $B_{min}$, $\eta$ between the measurement and the simulation is shown in Table V. The error of the magnetic field uniformity is 0.333%, which meets the requirements.

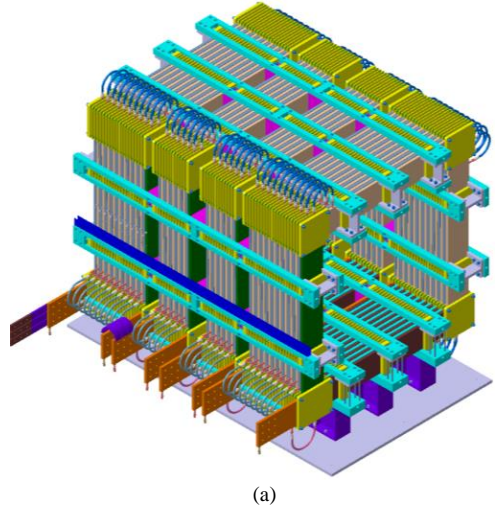

(a)

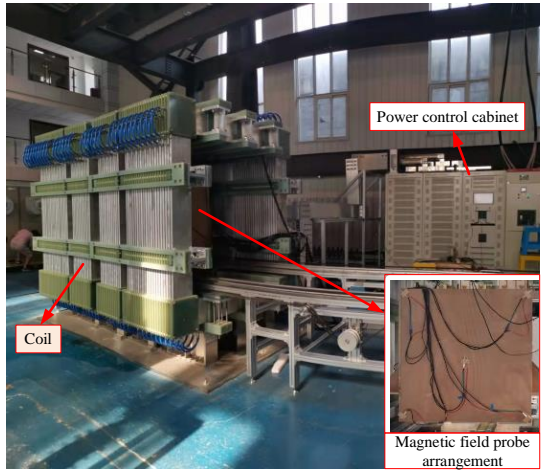

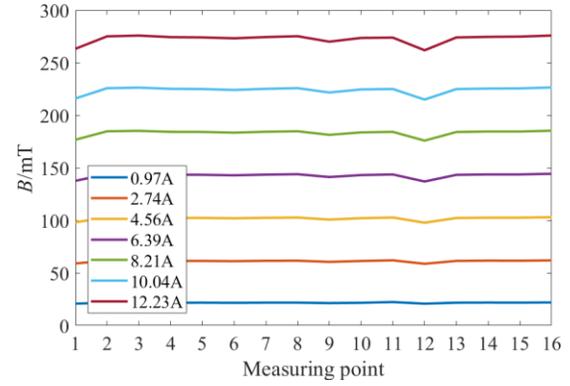

Fig. 11. Measured magnetic field distribution at different current levels.

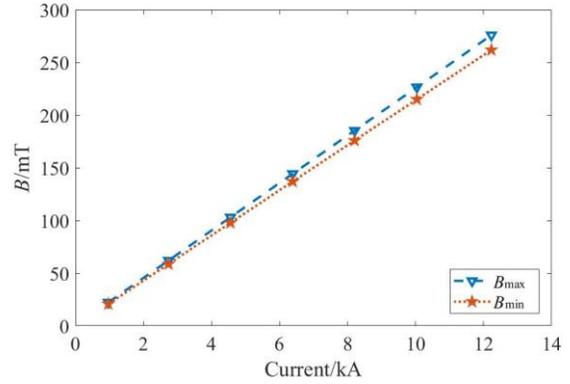

Fig. 10. Testing experimental: (a) Testing model of the four-coil system, (b)Test system experimental platform

TABLE IV
ERROR COMPARISON BETWEEN THE MEASUREMENT AND THE FEM

| Point | $B$/mT (FEM) | $B$/mT (Measurement) | Error |
|---|---|---|---|
| 1 | 261.07 | 263.26 | -0.84% |
| 2 | 272.81 | 275.11 | -0.84% |
| 3 | 272.79 | 275.87 | -1.13% |
| 4 | 272.76 | 274.38 | -0.59% |
| 5 | 272.76 | 274.10 | -0.49% |
| 6 | 268.42 | 273.28 | -1.81% |
| 7 | 269.53 | 274.42 | -1.81% |
| 8 | 272.00 | 275.24 | -1.19% |
| 9 | 274.24 | 270.05 | 1.53% |
| 10 | 275.11 | 273.62 | 0.54% |
| 11 | 274.47 | 273.91 | 0.20% |
| 12 | 261.02 | 261.86 | -0.32% |
| 13 | 272.82 | 274.07 | -0.46% |
| 14 | 272.80 | 274.62 | -0.67% |
| 15 | 272.79 | 274.86 | -0.76% |
| 16 | 272.79 | 275.90 | -1.14% |

TABLE V
ERROR COMPARISON THE MEASUREMENT AND THE FEM AT 12.23 KA

|  | $B_{max}$/mT | $B_{min}$/mT | $\eta$ |
|---|---|---|---|
| Analytic Method | 275.00 | 262.03 | 1.0495 |
| Measurement | 275.89 | 261.86 | 1.0530 |
| Error | -0.323% | 0.065% | 0.333% |

The measured magnetic field distribution at different current levels is shown in Fig. 11. It can be seen from the picture that point 1 and point 2 are the lowest points of the magnetic field at different current levels. Point 11-16 show relatively large magnetic field intensity. Fig. 12 shows that the current is almost linear with the maximum and minimum magnetic field value. Both Fig. 11 and Fig. 12 shows that the there is little difference between the maximum and minimum magnetic field at the test and the magnetic field uniformity is high.

Fig. 12. $B_{max}$, $B_{min}$ on different current level

## V. CONCLUSION

Based on ITER's high-uniformity magnetic field immunity testing system, the magnetic field calculation of four-coil configuration is studied. Firstly, the magnetic field calculation formulas of four-coil configuration under ideal and actual models are derived. The two main influencing factors of actual coil relative to ideal coil are considered, including helical structure and conductor cross-section. Then, a global parameter optimization method based on Lagrange Multiplier by KKT conditions is proposed to obtain the coil parameters of high-uniformity magnetic field. Finally, the correctness of the calculation is verified by the experiments of the existing equipment, and the coil structures with high-uniformity are compared by using the finite element method, which provides a theoretical basis for the subsequent actual manufacturing.

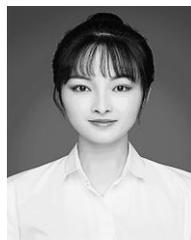

**Xi Deng** was born in Anhui, China, in 1995. She received the B.S. degree in Electrical Engineering and Automation from Anhui Agricultural University, Hefei, China. She is currently pursuing the Ph.D. degree in nuclear engineering with the Institute of Plasma Physics, Chinese Academy of Sciences, Hefei, China.

She majored in nuclear energy science and engineering. Her current research interests include fault diagnosis of high-power converter.

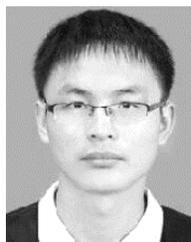

**Ya Huang** was born in Anhui, China, in 1990. He received the B.S. degree and the M.S. degree in electrical engineering from the Hefei University of Technology, Hefei, China, in 2011 and 2014, respectively, and the Ph.D. degree in the University of Science and Technology of China, Hefei, China, in 2021.

He majored in electronics, magnetic field analysis, and thermal analysis. His current research interests include EMC analysis and electric machine design.

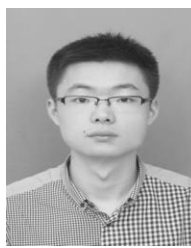

**Chenguang Wan** was born in Anhui, China, in 1995. He received a B.S. degree in mechanical design from Hefei University of Technology, Hefei, China. He is a Ph.D. candidate in plasma physics with the Institute of Plasma Physics, Chinese Academy of Sciences, Hefei, China.
His current research interests include tokamak modeling algorithm design and the intersection field in machine learning and modeling control

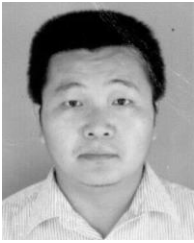

**Li Jiang** was born in Anhui, China, in 1981. He received the B.S. degree in electrical motor and automation from Anhui Polytechnic University, Wuhu, China, in 2005, and the Ph.D. degree in nuclear engineering from the Chinese Academy of Sciences, Hefei, China, in 2011.

He is currently an Associate Professor of the International Thermonuclear Experimental Reactor Project with the Institute of Plasma Physics, Chinese Academy of Sciences. He majored in electronics, magnetic field analysis, and integration design of high power supply system. His current research interests include power supply systems of fusion devices.

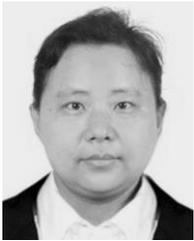

**Ge Gao** was born in Anhui, China, in 1975. She received the B.S. degree in electrical motor and automation from the Hefei University of Technology, Hefei, China, in 1996, and the Ph.D. degree in nuclear engineering from the Chinese Academy of Sciences, Hefei, in 2006.

She is currently a Professor with the International Thermonuclear Experimental Reactor Project, Institute of Plasma Physics, and Chinese Academy of Sciences. Her current research interest includes power supply systems of fusion devices.

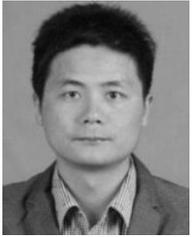

**Zhengyi Huang** was born in Anhui, China, in 1984. He received the B.S. degree in mechanical design manufacturing and automation from Anhui Polytechnic University and the M.S. degree in electrical engineering from the Hefei University of Technology, Hefei, China, in 2018. He is currently with the Institute of Plasma Physics, Chinese Academy of Sciences (ASIPP), Hefei. His current research interests include EMC analysis and DC busbar design

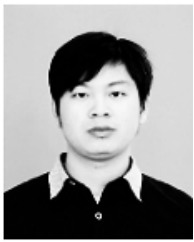

**Jie Zhang** was born in Anhui Province, China, in 1988. He received the B.S. degree in mechanical design and automation from Anhui Agricultural University, Hefei, China.

He is currently an Engineer with the Institute of Plasma Physics, Chinese Academy of Sciences, Hefei. His research interests include mechanical design and finite element analysis.